# High-efficiency grating couplers for pixel-level flat-top beam generation

Zhong-Tao Tian, Ze-Peng Zhuang, Zhi-Bin Fan, Xiao-Dong Chen[*] and Jian-Wen Dong

*School of Physics & State Key Laboratory of Optoelectronic Materials and Technologies, Sun Yat-sen University, Guangzhou 510275, China.*

[*]Corresponding author: chenxd67@mail.sysu.edu.cn

## ABSTRACT

We demonstrate a kind of grating coupler that generates high quality flat-top beam with a small beamwidth from photonic integrated circuits into free-space. The grating coupler is designed on a silicon-on-insulator wafer with a 220 nm-thick silicon layer and consists of a dual-etch grating (DG) and a distributed Bragg reflector (DBR). By adjusting the structural parameters of DG and DBR, a pixel-level (6.6 μm) flat-top beam with the vertical radiation of -0.5 dB and the mode match of 97% at 1550 nm is realized. Furthermore, a series of high-efficiency grating couplers are designed to create flat-top beam with different scales.

# 1. Introduction

Grating couplers are widely used in optical communication and sensing systems [1-9]. By converting the waveguide mode to the radiation mode, grating couplers have demonstrated a tremendous ability to couple the guided light into free space [10-13]. So far, several structures for high-efficiency fiber-chip coupling have been demonstrated, including an overlay structure [14,15], a Bragg or metallic mirror substrate [16-18], a slanted tilted waveguide [19,20], an asymmetry waveguide [21-23] or an apodized grating [24,25]. For example, by adding a poly-silicon layer deposition prior to grating etching, a highly efficient grating coupler can be obtained [14]. To reduce the transmission light and enhance optical coupling, a compact grating coupler combined with a reflector grating are proposed [26]. For the perfectly vertical coupling, a high efficiency two-layer vertical silicon grating coupler is optimized with the adjoint method and achieved a chip-to-fiber coupling efficiency of -0.035 dB at 1550 nm [21]. With the apodization technology, the grating coupler is designed to be apodized by varying the coupling strength for each grating period and a Gaussian intensity profile can be efficiently generated at the interface of grating [25]. On the other hand, highly efficient grating couplers have also been demonstrated with dual-etch structures [27-30]. Recently, a coupler design consisting of a two etch-step blazed and silicon nanopillars structure has been proposed to obtain a high efficiency and a low back reflection [31]. In addition, subwavelength grating metamaterial engineering has also been successfully adpoted to reduce index mismatch at structural transitions and improve coupling efficiency [32-36]. More recently, a silicon-based surface grating antenna design with a coupling efficiency of -0.5 dB and a directionality higher than 94% has been predicted by utilizing subwavelength-based L-shaped radiating elements in a 300-nm silicon core [37].

Most of the aforementioned grating couplers have been designed to efficiently generate the Gaussian intensity profile whose mode field diameter is around 10 μm [38]. This is suitable for certain applications such as fiber-chip coupling. For applications in the fluorescence imaging [39], optical phased arrays [40,41] and microscopy systems [42], a smaller grating coupler is desirable. There are some works on compact on-chip couplers [42-46]. For instance, a coupler design has been used as a component to couple light from a photonic chip waveguide to a microscopy system, focusing on applications of quantum-optics and bio-optics experiments [42]. Besides, for imaging applications, a pixel-level flat-top beam can provide a detailed analysis, increase the resolution and ensure more benefits [39,42]. Particularly, in fluorescence nanoscopy, the non-uniform illumination leads to position-dependent resolution and limits the field-of-view, which is harmful in high-quality super-resolution imaging [47]. On the other hand, for photodetectors, a smaller grating coupler can provide a compact waveguide integration scheme [48,49], and a flat-top beam with uniform intensity can alleviate the space charge effect [50]. Hence there is a need for compact and efficient couplers which can generate small flat-top beam.

In this letter, we present a silicon grating coupler that can transform the waveguide mode into the pixel-level flat-top mode. This is achieved exploiting a highly directional dual-etch grating (DG) and a distributed Bragg grating (DBR) in a 220-nm-thick silicon-on-insulator platform. The DG provides the high diffraction efficiency and low back reflection, and the DBR effectively increases the coupling length to enhance diffraction. The DBR can not only help to maintain high efficiency under a compact coupling length, but also improve the uniformity of radiation fields. By optimizing DG and DBR, the desired flat-top intensity and uniform wavefront in the orthogonal direction can be achieved. The grating coupler generates a pixel-level (6.6 μm) flat-top beam, and shows the vertical radiation of -0.5 dB and the mode match of 97%. Lastly, we demonstrate that this design is effective for grating couplers with various coupling lengths. This paper is organized as follows. The design concept and methodology of the proposed grating coupler are described in section 2. The simulation results are presented in section 3, the fabrication tolerance is analyzed in section 4, the grating couplers with different lengths are discussed in section 5, and the conclusions are summarized in section 6.

## 2. Design concept and methodology

In this section, we first explain the design principles of each part of the grating coupler, including DG, DBR and connecting waveguide. Then we specifically analyze the reason why DG can achieve high efficiency. The grating coupler comprises a DG, and a DBR in an SOI platform with a 220-nm-thick waveguide, 2-μm buried oxide and 2-μm silica cladding (Figure 1a). The DG implements an L-shaped structure with a shallow etch of 70 nm. Compared to the single-etch grating, DG enhances the diffraction strength due to its blazing effect [51]. In addition, the subwavelength pillars in DG provide the anti-reflection effect and reduce the back reflection [52], resulting in a high diffraction efficiency in the upward direction. The DBR redirects the forward transmission light and makes it passing through the DG twice. Therefore, the length of the grating coupler can be reduced while maintaining a high efficiency. Besides, the DBR also helps to reshape the radiation mode to match the flat-top mode. Without the DBR, the uniform DG will have an exponentially decaying light fields along the in-plane propagation direction. Compared with the flat-top mode, its optical intensity at the beginning of the DG is too strong and that at the end is too weak. The DBR can reflect back the forward transmission light and compensate the weak part of light. Thus, the diffraction fields can be reshaped to match the flat-top mode profile. Lastly, a silicon waveguide connects the DG and DBR, and its length has a strong impact on efficiency. Through the careful design of DG, DBR and connecting waveguide, this grating coupler can effectively obtain the pixel-level flat-top beam. In the following text, we will demonstrate the design principles and procedure of the grating coupler in detail. Throughout this work, we utilize a particle swarm optimization algorithm and 2D Finite-Difference Time-Domain simulations.

The unit cell of DG has two grooves with different etching depths, i.e., shallowly etched and fully etched units (right upper inset of Figure 1a). The theoretical model of DG has been discussed in previous studies [53], and we briefly summarize its design principle. Light propagating from left to right within the waveguide will encounter shallowly or fully etched grooves which act as a phased array of scatterers and couple light out of plane efficiently. As shown in Figure 1c, assuming that light is emitted from the center of the sidewall of the groove [54], the phase difference between light propagating from the fully etched grating and the shallowly etched grating can be decomposed into a vertical phase difference and a lateral phase difference. When the vertical and lateral phase difference accumulate $\pi/2$ phase shifts respectively, constructive interference in the upward direction and destructive interference in the downward direction can be realized, and then unidirectional radiation can be achieved.

## 3. Results and discussion

In this section, we design and investigate a grating coupler, which consists of a 10 periods DG, a DBR and a connecting waveguide. Firstly, the DG is optimized to achieve the highest vertical radiation at the wavelength of 1550 nm. As shown in Figure 1b, the light incident to DG has four possible output channels, namely vertical radiation, back re-flection, bottom leakage and forward transmission. The vertical radiation is defined as $\Gamma = P_{vertical}/P_{input}$ where $P_{vertical}$ is the vertically radiated power and $P_{input}$ is the incident power. In the simulation, a waveguide mode with transverse electric polarization is incident from the left. The structural parameters of $d_1$, $d_2$, $d_3$, $d_4$, $d_5$ are optimized using the particle swarm algorithm which has been widely used in various optical designs such as gratings [31], lens [55], nanophotonic structures [56]. The detailed algorithm implementation can refer to article [57]. We assume that the DG is covered by $SiO_2$ as a top-cladding layer and the refractive indices of Si and $SiO_2$ are 3.48 and 1.44, respectively. The obtained optimal structural parameters are: $d_1$ = 92 nm, $d_2$ = 84 nm, $d_3$ = 252 nm, $d_4$ = 133 nm and $d_5$ = 99 nm. The total length of DG with 10 periods is 6.6 μm.

The above DG enhances the vertical radiation and suppresses the bottom leakage. This is useful for improving the efficiency. However, when reducing the dimension of DG, the light cannot be radiated out-of-plane at one time and there will be a lot of forward transmission light. To deal with this issue, we adopt the compensation effect of the DBR. The DBR is designed based on the Bragg condition of $\beta_m = \beta + mG$, where $\beta_m$ represents the mth diffraction wave vector, $\beta$ is the incident light wave vector, and $mG$ is the wave vector introduced by the grating (Figure 2a). The DBR uses the negative first-order diffraction (i.e. $m$ = -1) to reflect the forward transmission light. So in this case, the Bragg condition can be written as:

$$n_{core}\Lambda_r ff + n_{clad}\Lambda_r(1-ff) = \frac{\lambda}{2}, \qquad (1)$$

where $n_{core}$ and $n_{clad}$ are the effective refractive indices of Si core and $SiO_2$ cladding, $\Lambda_r$ is the period length of reflection grating, $ff$ is the fill factor ($ff$ = 0.5) and $\lambda$ is the center wavelength ($\lambda$ = 1550 nm). By substituting these values into the Bragg condition, the period length is calculated to be around 322 nm. To accurately find the parameters corresponding to the highest reflectivity, we simulate the DBR with a period length from 200 nm to 500 nm. In the simulation, both the wavelength interval and the period length interval are 1 nm. The results are shown in Figure 2b. We selected the period length to be 311 nm to achieve high reflectivity in a wide spectral range. Then, the designed DBR is placed after the DG to form a grating coupler. To see the compensation effect of DBR, we compare the forward transmission light and vertical radiation of the grating coupler with and without DBR (Figure 2c,d). The forward transmission is greatly suppressed since it drops from about -5 dB to about -30 dB. As a result, the vertical radiation is enhanced near 1550 nm.

Note that the spacing between DG and DBR is another important parameter because it significantly affects the coupling efficiency. Figure 3 plots the vertical radiation, back reflection and bottom leakage as a function of the connecting waveguide width $W_d$. They change periodically with the increase of $W_d$. At some points with the highest vertical radiation, the back reflection and the bottom leakage are suppressed. In our paper, the connecting waveguide length is chosen to be $W_d = 90$ nm. In this case, the light reflected by the DBR destructively interferes with the light reflected by the DG. Similarly, destructive interference occurs by emitting bottom-coupled light from the left and another emitting bottom-coupled light reflected by the DBR and from the right.

Overall, a high efficiency grating coupler is obtained by carefully selecting the structural parameters of DG, DBR and the connecting waveguide. The simulated electric fields of the optimized grating coupler are shown in Figure 4a. Most of the light is vertically coupled along the $+y$ direction, and electric fields in other directions are weak, indicating that the unwanted diffraction loss is suppressed. In our optimized structure, the vertical radiation reaches -0.5 dB. Besides of the diffraction efficiency, we also analyze the mode matching capability of the grating coupler. To do this, the $E_z$ slices of the optimized beam is shown in Figure 4b,c. Obviously, both the amplitude and phase of the simulated electric field are close to the flat-top contour of the design target over most of the width of the beam, only slightly offset near the edge of the beam. This offset does not lead to a significant reduction in mode match. To quantitatively measure the mode match between simulation and target flat-top mode, we define a mode match coefficient based on the overlap integral [3]:

$$\eta = \left|\frac{\int (E_s(x) \cdot E_t^*(x))^2 dx}{\int E_s^2(x)dx \cdot \int E_t^2(x)dx}\right|. \tag{2}$$

Here, $E_s(x)$ and $E_t(x)$ respectively correspond to the electric fields of the simulation wave and target mode (represented by the super Gaussian function $E_t(x) = exp[-(x/w)^N]$, where $w$ is the half-maximum beam width and $N = 24$). Ideally, the out-coupled mode should exhibit a maximum overlap with the flat- top mode, i.e., $\eta = 100\%$. In our optimized structure, the mode match exceeds 97% at the wavelength of 1550 nm and is close to the ideal result. Besides, the mode match is larger than 80% (i.e. 1 dB) in the whole C band, and shows broad spectral characteristics (Figure 4d). Moreover, the far-field radiations of light with different wavelengths are calculated to see the diffraction directionality (Figure 4e). The designed grating coupler can keep vertical diffraction well at 1550 nm. The radiation angle wavelength shift is 0.11°/nm over 1500 nm - 1600 nm.

## 4. Fabrication tolerance analysis

The experimental realization of the above grating coupler is feasible. The proposed structure relies on a simple double etching process and is compatible with standard silicon-on-insulator technology. Its minimum feature size is larger than 80 nm, which ensures a convenient platform for further fabrication. These results are promising for current and future applications, especially considering that this structure has been fabricated using the existing CMOS process [31]. Since there might be fabrication errors in the grating width, the thickness of the shallowly etched area, and the connecting waveguide length during the fabrication, vertical radiation and mode match are calculated for cases where fabrication errors happen. Here, the errors of the grating width caused by the full etching and the shallow etching are assumed to be $\Delta d_F$ and $\Delta d_S$. While the errors of the length of the connecting waveguide and the shallowly etched depth are defined as $\Delta d_L$ and $\Delta d_T$, respectively. Thus, these fabrication errors are expressed by: $d_1 = d_{1,0} + \Delta d_F$, $d_2 = d_{2,0} - \Delta d_F$, $d_3 = d_{3,0} + \Delta d_F/2 - \Delta d_S/2$, $d_4 = d_{4,0} + \Delta d_F/2 + \Delta d_S/2$, $d_5 = d_{5,0} - \Delta d_F$, $d_L = d_{L,0} - \Delta d_L$, $d_T = d_{T,0} - \Delta d_T$, where parameters with subscript "0" represent the ideal values. As shown in Figure 5a, vertical radiation and mode match show relatively large fluctuations as the width changes. The performance of the grating coupler is more affected by the fully etched area. Furthermore, the performance variations remain comparatively small for the fabrication error of shallowly etched area (Figure 5b). We also consider the impact that a variation of the waveguide length could have on the grating performance. This variability is much smaller than the fabrication variations related to the lithography. Figure 5c shows that the performance of the grating coupler is slightly affected within the selected range of waveguide length variations. Figure 5d shows the calculation result of the change in vertical radiation and mode match with respect to the depth of the shallowly etched area. The vertical radiation of > 3 dB and mode match of > 80% can be maintained for deviations of < ±15 nm.

**5. Grating couplers with different lengths**

In the above, we have designed a grating coupler with pixel-level of 6.6 μm. In this section, we verify this design method can be applied to grating couplers with other pixel-level lengths. In Figure 6a, we calculate the vertical radiation and mode match of DG with 1 to 21 periods and all other structural parameters are the same as those in Figure 4. The vertical radiation and mode match are high for when the number of periods is larger than 6, but the vertical radiation is low when the number of periods is smaller than 6. As an example, the $E_z$ slice of output modes of DG with 3 periods shows that although most of the light is radiated upward, the electric fields at the bottom are also strong. This indicates that there exists the high bottom leakage and the DG with small periods should be further modified to achieve the high vertical radiation. Therefore, we use the particle swarm algorithm to find the optimal structural parameters of DGs with 1 to 6 periods. The optimized DGs yields a grating coupler with a higher vertical radiation than that of the unoptimized one (Figure 6b). Meanwhile, the mode match of the optimized grating coupler remained almost unchanged (~ 95%).

## 6. Conclusions

In conclusion, a grating coupler that efficiently couples a fundamental mode of silicon waveguide to a pixel-level flat-top beam in free space is designed and numerically demonstrated. Several different pixel-level grating couplers are verified as well. Specifically, the DG offers a high-efficiency diffraction and the followed DBR provides compensation effect and makes sure reducing the circuit footprint. Meanwhile, the flat-top beam is achieved by optimizing the structural parameters of the grating coupler. The vertical radiation of the grating coupler is above -0.5 dB and its mode match is up to 97%. This kind of grating coupler can be a module for the interaction of the photonic chip with other systems like biosensing, high-precision image sensors, structured light illumination and so on.

**Acknowledgments**

This work was supported by State Key Research Development Program of China (No. 2019YFB2203502), National Natural Science Foundation of China (Nos. 12074443, and 62035016), Guangdong Basic and Applied Basic Research Foundation (No. 2019B151502036), Guangzhou Science, Technology and Innovation Commission (Grant No. 202102020693), Fundamental Research Funds for the Central Universities (Grant Nos. 20lgzd29 and 20lgjc05).

# REFERENCES

1. Taillaert, D.; Bienstman, P.; Baets, R. Compact efficient broadband grating coupler for silicon-on-insulator waveguides. Opt Lett **2004**, 29, 2749-2751.
2. Zaoui, W.S.; Kunze, A.; Vogel, W.; Berroth, M.; Butschke, J.; Letzkus, F.; Burghartz, J. Bridging the gap between optical fibers and silicon photonic integrated circuits. Opt Express **2014**, 22, 1277-1286.
3. Xu, X.C.; Subbaraman, H.; Covey, J.; Kwong, D.; Hosseini, A.; Chen, R.T. Complementary metal-oxide-semiconductor compatible high efficiency subwavelength grating couplers for silicon integrated photonics. Appl Phys Lett **2012**, 101.
4. He, L.; Liu, Y.; Galland, C.; Lim, A.E.J.; Lo, G.Q.; Baehr-Jones, T.; Hochberg, M. A High-Efficiency Nonuniform Grating Coupler Realized With 248-nm Optical Lithography. Ieee Photonic Tech L **2013**, 25, 1358-1361.
5. Song, J.H.; Doany, F.E.; Medhin, A.K.; Dupuis, N.; Lee, B.G.; Libsch, F.R. Polarization-independent nonuniform grating couplers on silicon-on-insulator. Opt Lett **2015**, 40, 3941-3944.
6. Marchetti, R.; Lacava, C.; Carroll, L.; Gradkowski, K.; Minzioni, P. Coupling strategies for silicon photonics integrated chips. Photonics Research **2019**, 7, 201-239.
7. Melati, D.; Grinberg, Y.; Dezfouli, M.K.; Janz, S.; Cheben, P.; Schmid, J.H.; Sanchez-Postigo, A.; Xu, D.X. Mapping the global design space of nanophotonic components using machine learning pattern recognition. Nat Commun **2019**, 10.
8. Cheng, L.; Mao, S.; Li, Z.; Han, Y.; Fu, H.Y. Grating couplers on silicon photonics: Design principles, emerging trends and practical issues. Micromachines **2020**, 11, 666.
9. Son, G.; Han, S.; Park, J.; Kwon, K.; Yu, K. High-efficiency broadband light coupling between optical fibers and photonic integrated circuits. Nanophotonics **2018**, 7, 1845-1864.
10. Kim, S.; Westly, D.A.; Roxworthy, B.J.; Li, Q.; Yulaev, A.; Srinivasan, K.; Aksyuk, V.A. Photonic waveguide to free-space Gaussian beam extreme mode converter. Light: Science & Applications **2018**, 7, 72.
11. Ropp, C.; Yulaev, A.; Westly, D.; Simelgor, G.; Aksyuk, V. Meta-grating outcouplers for optimized beam shaping in the visible. Opt Express **2021**, 29, 14789-14798.
12. Romero, J.; Giovannini, D.; Franke-Arnold, S.; Barnett, S.M.; Padgett, M.J. Increasing the dimension in high-dimensional two-photon orbital angular momentum entanglement. Physical Review A **2012**, 86, 012334.
13. Ropp, C.; Yulaev, A.; Zhu, W.; Westly, D.A.; Simelgor, G.; Agrawal, A.; Papp, S.; Aksyuk, V. Multi-Beam Integration for On-chip Quantum Devices. In Proceedings of the Conference on Lasers and Electro-Optics, San Jose, California, 2021/05/09, **2021**; p. STh4A.7.
14. Roelkens, G.; Van Thourhout, D.; Baets, R. High efficiency Silicon-on-Insulator grating coupler based on a poly-Silicon overlay. Opt Express **2006**, 14, 11622-11630.
15. Vermeulen, D.; Selvaraja, S.; Verheyen, P.; Lepage, G.; Bogaerts, W.; Absil, P.; Van Thourhout, D.; Roelkens, G. High-efficiency fiber-to-chip grating couplers realized using an advanced CMOS-compatible Silicon-On-Insulator platform. Opt Express **2010**, 18, 18278-18283.
16. Romero-Garcia, S.; Merget, F.; Zhong, F.; Finkelstein, H.; Witzens, J. Visible wavelength silicon nitride focusing grating coupler with AlCu/TiN reflector. Opt Lett **2013**, 38, 2521-2523.
17. Luo, Y.N.; Nong, Z.C.; Gao, S.Q.; Huang, H.M.; Zhu, Y.T.; Liu, L.; Zhou, L.D.; Xu, J.; Liu, L.; Yu, S.Y.; et al. Low-loss two-dimensional silicon photonic grating coupler with a backside metal mirror.

Opt Lett **2018**, 43, 474-477.
18. Zaoui, W.S.; Kunze, A.; Vogel, W.; Berroth, M. CMOS-Compatible Polarization Splitting Grating Couplers With a Backside Metal Mirror. Ieee Photonic Tech L **2013**, 25, 1395-1397.
19. Schrauwen, J.; Van Laere, F.; Van Thourhout, D.; Baets, R. Focused-ion-beam fabrication of slanted grating couplers in silicon-on-insulator waveguides. Ieee Photonic Tech L **2007**, 19, 816-818.
20. Bin, W.; Jianhua, J.; Nordin, G.P. Embedded slanted grating for vertical coupling between fibers and silicon-on-insulator planar waveguides. Ieee Photonic Tech L **2005**, 17, 1884-1886.
21. Michaels, A.; Yablonovitch, E. Inverse design of near unity efficiency perfectly vertical grating couplers. Opt Express **2018**, 26, 4766-4779.
22. Dai, M.; Ma, L.; Xu, Y.; Lu, M.; Liu, X.; Chen, Y. Highly efficient and perfectly vertical chip-to-fiber dual-layer grating coupler. Opt Express **2015**, 23, 1691-1698.
23. Su, L.; Trivedi, R.; Sapra, N.V.; Piggott, A.Y.; Vercruysse, D.; Vučković, J. Fully-automated optimization of grating couplers. Opt Express **2018**, 26, 4023-4034.
24. Mossberg, T.W.; Greiner, C.; Iazikov, D. Interferometric amplitude apodization of integrated gratings. Opt Express **2005**, 13, 2419-2426.
25. Marchetti, R.; Lacava, C.; Khokhar, A.; Chen, X.; Cristiani, I.; Richardson, D.J.; Reed, G.T.; Petropoulos, P.; Minzioni, P. High-efficiency grating-couplers: demonstration of a new design strategy. Sci Rep-Uk **2017**, 7.
26. Taillaert, D.; Bogaerts, W.; Bienstman, P.; Krauss, T.F.; Daele, P.V.; Moerman, I.; Verstuyft, S.; Mesel, K.D.; Baets, R. An out-of-plane grating coupler for efficient butt-coupling between compact planar waveguides and single-mode fibers. IEEE Journal of Quantum Electronics **2002**, 38, 949-955.
27. Alonso-Ramos, C.; Cheben, P.; Ortega-Monux, A.; Schmid, J.H.; Xu, D.X.; Molina-Fernandez, I. Fiber-chip grating coupler based on interleaved trenches with directionality exceeding 95%. Opt Lett **2014**, 39, 5351-5354.
28. Benedikovic, D.; Alonso-Ramos, C.; Guerber, S.; Le Roux, X.; Cheben, P.; Dupré, C.; Szelag, B.; Fowler, D.; Cassan, É.; Marris-Morini, D.; et al. Sub-decibel silicon grating couplers based on L-shaped waveguides and engineered subwavelength metamaterials. Opt Express **2019**, 27, 26239-26250.
29. Chen, X.; Thomson, D.J.; Crudginton, L.; Khokhar, A.Z.; Reed, G.T. Dual-etch apodised grating couplers for efficient fibre-chip coupling near 1310 nm wavelength. Opt Express **2017**, 25, 17864-17871.
30. Watanabe, T.; Fedoryshyn, Y.; Leuthold, J. 2-D grating couplers for vertical fiber coupling in two polarizations. IEEE Photonics Journal **2019**, 11, 1-9.
31. Watanabe, T.; Ayata, M.; Koch, U.; Fedoryshyn, Y.; Leuthold, J. Perpendicular Grating Coupler Based on a Blazed Antiback-Reflection Structure. J Lightwave Technol **2017**, 35, 4663-4669.
32. Zhou, W.; Cheng, Z.; Chen, X.; Xu, K.; Sun, X.; Tsang, H. Subwavelength engineering in silicon photonic devices. Ieee J Sel Top Quant **2019**, 25, 1-13.
33. Benedikovic, D.; Alonso-Ramos, C.; Perez-Galacho, D.; Guerber, S.; Vakarin, V.; Marcaud, G.; Le Roux, X.; Cassan, E.; Marris-Morini, D.; Cheben, P.; et al. L-shaped fiber-chip grating couplers with high directionality and low reflectivity fabricated with deep-UV lithography. Opt Lett **2017**, 42, 3439-3442.
34. Dezfouli, M.K.; Grinberg, Y.; Melati, D.; Cheben, P.; Schmid, J.H.; Sánchez-Postigo, A.; Ortega-Moñux, A.; Wangüemert-Pérez, G.; Cheriton, R.; Janz, S. Perfectly vertical surface grating couplers using subwavelength engineering for increased feature sizes. Opt Lett **2020**, 45, 3701-3704.
35. Halir, R.; Cheben, P.; Schmid, J.H.; Ma, R.; Bedard, D.; Janz, S.; Xu, D.X.; Densmore, A.;

Lapointe, J.; Molina-Fernandez, I. Continuously apodized fiber-to-chip surface grating coupler with refractive index engineered subwavelength structure. Opt Lett **2010**, 35, 3243-3245.
36. Cheben, P.; Halir, R.; Schmid, J.H.; Atwater, H.A.; Smith, D.R. Subwavelength integrated photonics. Nature **2018**, 560, 565-572.
37. Khajavi, S.; Melati, D.; Cheben, P.; Schmid, J.H.; Liu, Q.; Xu, D.X.; Winnie, N.Y. Compact and highly-efficient broadband surface grating antenna on a silicon platform. Opt Express **2021**, 29, 7003-7014.
38. Chen, X.; Li, C.; Fung, C.K.Y.; Lo, S.M.G.; Tsang, H.K. Apodized Waveguide Grating Couplers for Efficient Coupling to Optical Fibers. Ieee Photonic Tech L **2010**, 22, 1156-1158.
39. Khaw, I.; Croop, B.; Tang, J.L.; Mohl, A.; Fuchs, U.; Han, K.Y. Flat-field illumination for quantitative fluorescence imaging. Opt Express **2018**, 26, 15276-15288.
40. Sun, J.; Timurdogan, E.; Yaacobi, A.; Hosseini, E.S.; Watts, M.R. Large-scale nanophotonic phased array. Nature **2013**, 493, 195-199.
41. Pita, J.L.; Aldaya, I.; Dainese, P.; Hernandez-Figueroa, H.E.; Gabrielli, L.H. Design of a compact CMOS-compatible photonic antenna by topological optimization. Opt Express **2018**, 26, 2435-2442.
42. Zhu, Y.P.; Wang, J.; Xie, W.Q.; Tian, B.; Li, Y.L.; Brainis, E.; Jiao, Y.Q.; Van Thourhout, D. Ultra-compact silicon nitride grating coupler for microscopy systems. Opt Express **2017**, 25, 33297-33304.
43. Luxmoore, I.J.; Wasley, N.A.; Ramsay, A.J.; Thijssen, A.C.T.; Oulton, R.; Hugues, M.; Kasture, S.; Achanta, V.G.; Fox, A.M.; Skolnick, M.S. Interfacing spins in an InGaAs quantum dot to a semiconductor waveguide circuit using emitted photons. Physical review letters **2013**, 110, 037402.
44. Shimizu, W.; Nagai, N.; Kohno, K.; Hirakawa, K.; Nomura, M. Waveguide coupled air-slot photonic crystal nanocavity for optomechanics. Opt Express **2013**, 21, 21961-21969.
45. Faraon, A.; Fushman, I.; Englund, D.; Stoltz, N.; Petroff, P.; Vučković, J. Dipole induced transparency in waveguide coupled photonic crystal cavities. Opt Express **2008**, 16, 12154-12162.
46. Sánchez-Postigo, A.; Ortega-Moñux, A.; Pereira-Martín, D.; Molina-Fernández, Í.; Halir, R.; Cheben, P.; Penadés, J.S.; Nedeljkovic, M.; Mashanovich, G.Z.; Wangüemert-Pérez, J.G. Design of a suspended germanium micro-antenna for efficient fiber-chip coupling in the long-wavelength mid-infrared range. Opt Express **2019**, 27, 22302-22315.
47. Douglass, K.M.; Sieben, C.; Archetti, A.; Lambert, A.; Manley, S. Super-resolution imaging of multiple cells by optimized flat-field epi-illumination. Nat Photonics **2016**, 10, 705-708.
48. Li, H.; Liu, Y.; Miao, C.; Zhang, M.; Zhou, W.; Tang, C.; Li, E. High-performance binary blazed grating coupler used in silicon-based hybrid photodetector integration. Optical Engineering **2014**, 53, 097106.
49. Roelkens, G.; Brouckaert, J.; Taillaert, D.; Dumon, P.; Bogaerts, W.; Van Thourhout, D.; Baets, R.; Nötzel, R.; Smit, M. Integration of InP/InGaAsP photodetectors onto silicon-on-insulator waveguide circuits. Opt Express **2005**, 13, 10102-10108.
50. Zuo, Y.; Yu, Y.; Zhang, Y.; Zhou, D.; Zhang, X. Integrated high-power germanium photodetectors assisted by light field manipulation. Opt Lett **2019**, 44, 3338-3341.
51. Benedikovic, D.; Alonso-Ramos, C.; Cheben, P.; Schmid, J.H.; Wang, S.R.; Xu, D.X.; Lapointe, J.; Janz, S.; Halir, R.; Ortega-Monux, A.; et al. High-directionality fiber-chip grating coupler with interleaved trenches and subwavelength index-matching structure. Opt Lett **2015**, 40, 4190-4193.
52. Wang, Y.; Wang, X.; Flueckiger, J.; Yun, H.; Shi, W.; Bojko, R.; Jaeger, N.A.F.; Chrostowski, L. Focusing sub-wavelength grating couplers with low back reflections for rapid prototyping of silicon photonic circuits. Opt Express **2014**, 22, 20652-20662.

53. Flory, C.A. Analysis of directional grating-coupled radiation in waveguide structures. IEEE Journal of Quantum Electronics **2004**, 40, 949-957.

54. Chen, B.; Li, Y.; Zhang, L.; Li, Y.; Liu, X.; Tao, M.; Hou, Y.; Tang, H.; Zhi, Z.; Gao, F.; et al. Unidirectional large-scale waveguide grating with uniform radiation for optical phased array. Opt Express **2021**, 29, 20995-21010.

55. Duan, F.; Zhu, W.-L.; Han, Y.; Ju, B.-F.; Beaucamp, A. Chromatically multi-focal optics based on micro-lens array design. Opt Express **2020**, 28, 24123-24135.

56. Digani, J.; Hon, P.W.C.; Davoyan, A.R. Framework for Expediting Discovery of Optimal Solutions with Blackbox Algorithms in Non-Topology Photonic Inverse Design. ACS Photonics 2022.

57. Wang, D.; Tan, D.; Liu, L. Particle swarm optimization algorithm: an overview. Soft Computing **2018**, 22, 387-408.

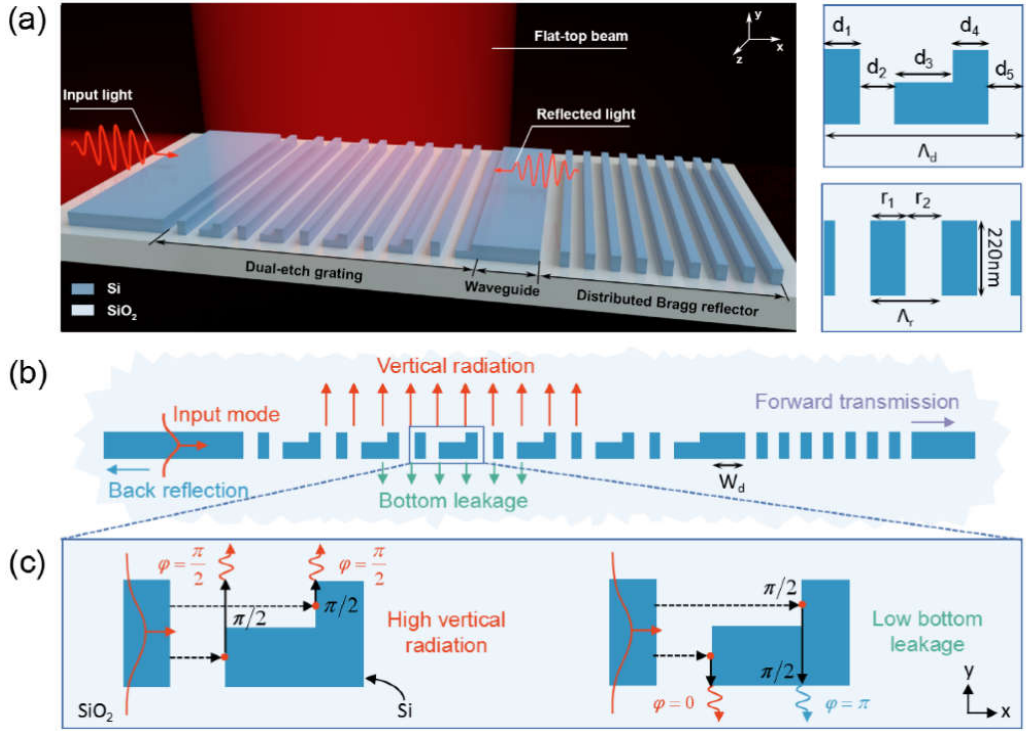

**FIG. 1.** (a) Schematic of the grating coupler consisting of DG and DBR. The DG utilizes a blazed sub-wavelength structure with period and optimization variables [$d_1$, $d_2$, $d_3$, $d_4$, $d_5$]. The period length of the DBR is $\Lambda_r$. (b) Description of the operation of the grating coupler. Light is incident from the left waveguide and has four output channels, i.e., the channel along the vertical radiation, back refection, bottom leakage and forward transmission. (c) Description of the operation principle of DG.

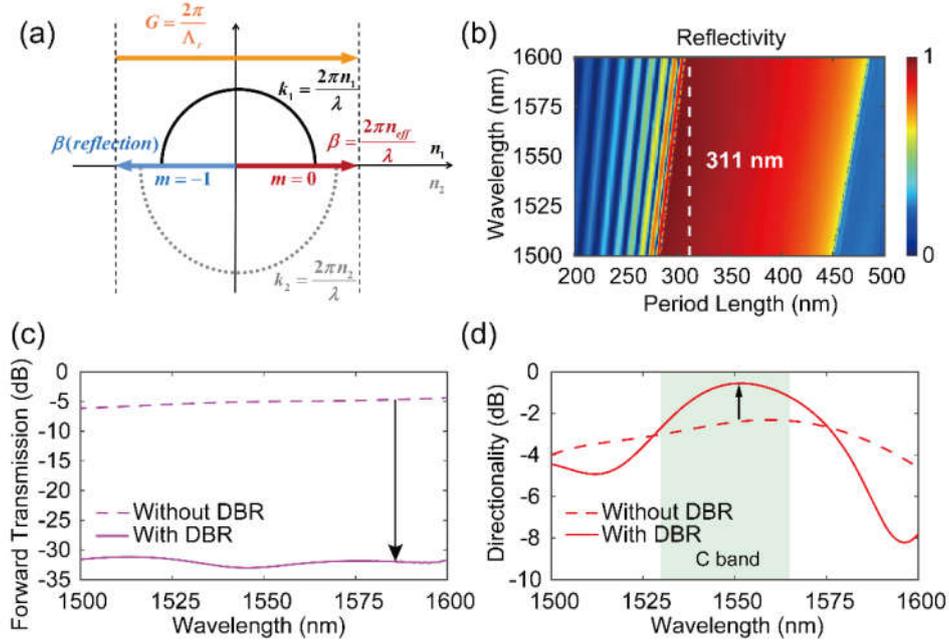

**FIG. 2.** (a) Wave-vector diagram of the DBR which uses the negative first-order diffraction to reflect the forward transmission light. (b) Reflectivity phase diagram of the DBR whose fill factor is 0.5. (c, d) Comparison of the forward transmission and vertical radiation for structures without and with DBR.

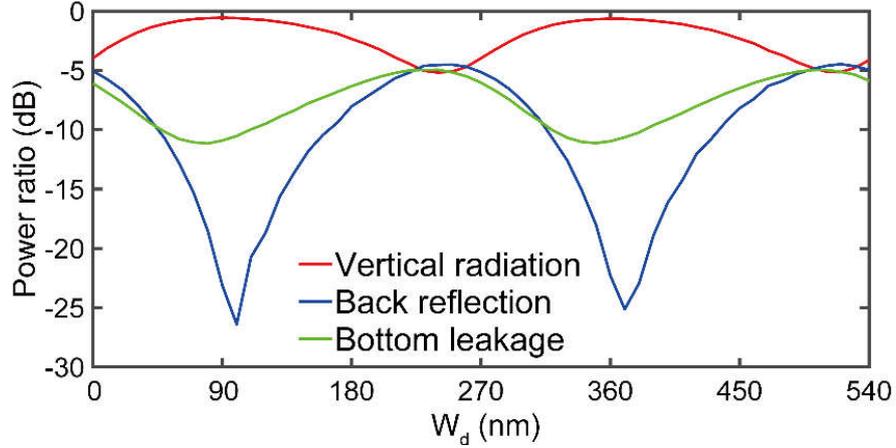

**FIG. 3. Vertical radiation (red), back reflection (blue) and bottom leakage (green) as a function of the connecting waveguide width.**

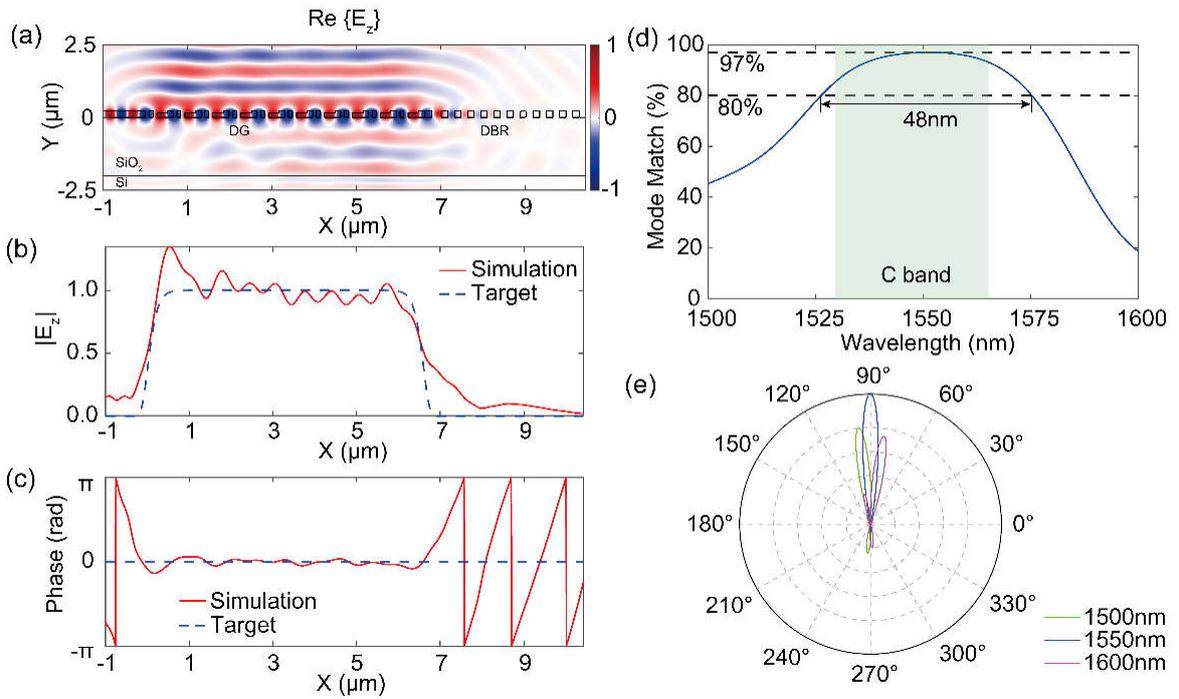

**FIG. 4. Performance of the grating coupler with a 10 periods DG and a DBR.** (a) The real part of Ez shows the simulated radiation with the high vertical radiation and flat wavefront. (b, c) Comparison between the simulated (b) amplitude and (c) phase of Ez for the optimized grating coupler and the target flat-top wavefront. (d) The mode match results of the designed structure. (e) The far-field radiation performance of the optimized grating coupler.

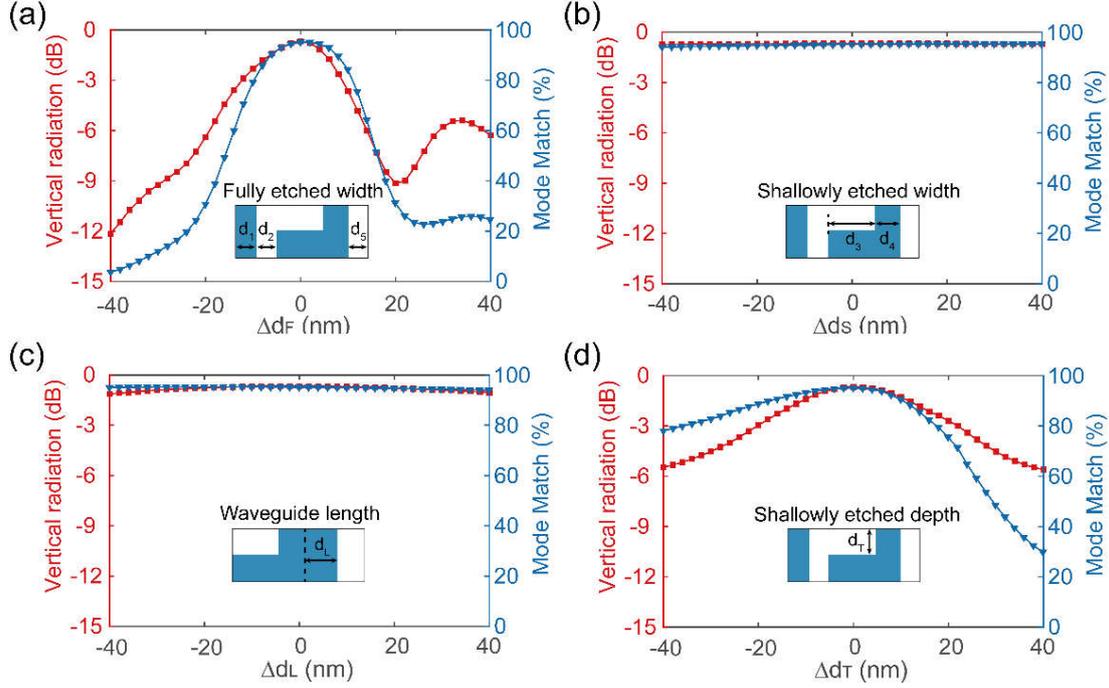

**FIG. 5. Fabrication tolerances of the vertical radiation and mode match.** (a) Width error caused by full etching process. (b) Width error caused by shallow etching process. (c) Length error of connecting waveguide. (d) Depth error caused by shallow etching process.

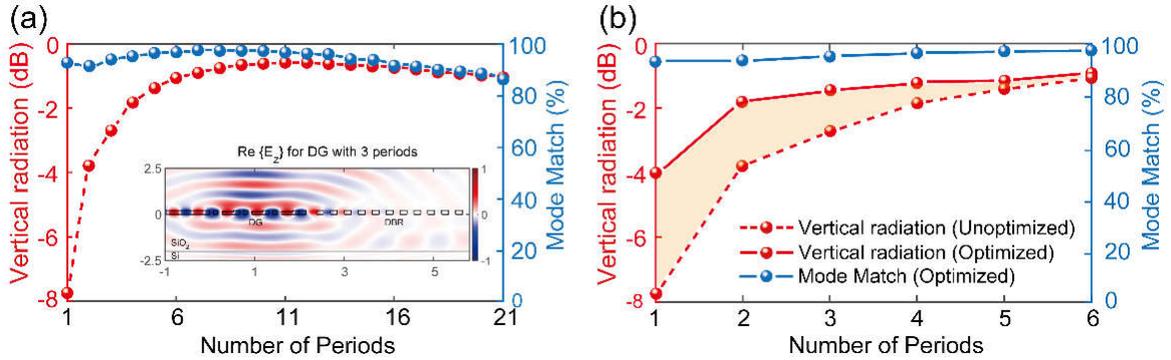

**FIG. 6. Performance of DGs with different numbers of periods.** (a) The vertical radiation is deteriorated in DGs with small periods. The inset shows the real part of Ez of the DG with 3 periods. Here, structural parameters except the number of periods of DG are the same as those in Figure 4. (b) The optimal performance of DGs (solid line) with 1 to 6 periods whose structural parameters are found by using the particle swarm algorithm to ensure low diffraction loss.